# Effect of Cobalt Content on the Electrochemical Properties and Structural Stability of NCA Type Cathode Materials

Kamalika Ghatak[1], Swastik Basu[2], Tridip Das[3], Hemant Kumar[4], Dibakar Datta[1,*]

[1] Department of Mechanical and Industrial Engineering, Newark College of Engineering, New Jersey Institute of Technology (NJIT), Newark, NJ 07102, USA

[2] Department of Mechanical, Aerospace, and Nuclear Engineering, Rensselaer Polytechnic Institute, Troy, NY 12180, USA

[3] Department of Chemical Engineering and Materials Science, Michigan State University, East Lansing, MI 48824, USA

[4] Department of Materials Science and Engineering, University of Pennsylvania, Philadelphia, PA 19104, USA

**Corresponding Author:**

Dibakar Datta; Phone: 973-596-3647; Email: dibakar.datta@njit.edu

**Abstract**

At present, the most common type of cathode materials, NCA ($Li_{1-x}Ni_{0.80}Co_{0.15}Al_{0.05}O_2$, $x$ = 0 to 1), have a very high concentration of cobalt. Since cobalt is toxic and expensive, the existing design of cathode materials is neither cost-effective nor environmentally benign. We have performed density functional theory (DFT) calculations to investigate electrochemical, electronic, and structural properties of four types of NCA cathode materials with the simultaneous decrease in Co content along with the increase in Ni content. Our results show that even if the cobalt concentration is significantly decreased from 16.70 % (NCA_I) to 4.20 % (NCA_IV), variation in intercalation potential and specific capacity is not significant. For example, in case of 50% Li concentration, the voltage drop is only ~17% while the change in specific capacity is negligible. Moreover, we have also explored the influence of sodium doping in the intercalation site on the electrochemical, electronic, and structural properties. By considering two extreme cases of NCAs (i.e., with highest and lowest Co content: NCA_I and NCA_IV respectively), we have demonstrated the importance of Na doping from the structural and electronic point of view. Our results provide insight into the design of environmentally benign, low-cost cathode materials with reduced cobalt concentration.

**Keywords:** Lithium-ion Batteries, NCA Cathode, Cobalt Content, Density Functional Theory, Electrochemical Properties, Structural Stability

1. **Introduction:**

One of the important requirements of cathode materials in lithium-ion batteries (LIBs) is their ability to intercalate lithium (Li) ions reversibly without causing significant changes to the atomic structure. The commonly used cathode materials are lithium Metal Oxides, $LiMO_2$, where the M represents a pure or a combination of transition metals such as cobalt (Co), nickel (Ni), manganese (Mn), iron (Fe), and titanium (Ti).[1] Among the transition metal oxides, lithium-cobalt-oxide, $LiCoO_2$ (LCO) is one of the well-studied cathode materials due to its excellent intercalation nature, and its first successful commercial usage was in the year 1980.[2] Other essential properties responsible for its commercial success are [3-4] low self-discharge, higher discharge voltage, around 500-700 deep discharge cycle life, and relatively high theoretical specific capacity as well as volumetric capacity. Some of the drawbacks associated with LCO are its low thermal stability, rapid capacity loss with cycling, and the presence of Co,[5] which makes it not only an expensive material but also very toxic. Among all the available cathode materials, LCO has the lowest reported thermal stability,[6] which is also the prime cause of adverse safety issues related to LCOs. For example, low thermal stability causes the exothermic release of oxygen when heated over a certain point and eventually releases large amounts of energy that leads to the explosion.[7] Therefore, in spite of the advantages, researchers are still looking for alternatives to LCO due to these drawbacks.[8]

Lithium-nickel-oxide, $LiNiO_2$ (or LNO) is one of the commercially available alternatives and has a layered structure similar to that of LCO as shown in **Figure 1.**[9] In LCO and LNO, the $Co^{3+}$ or $Ni^{3+}$ ions occupy $O^h$ sites adjoining neighboring $O^{2-}$ layers, respectively. The transition metal ions occupy alternate layers of octahedral sites and rests of the layers are filled with $Li^+$ ions. The overall crystal structure gives rise to rhombohedral geometry. The main reason behind the extensive commercial research on LNO remains the low production cost and low toxicity, compared to LCO as Ni is benign and much cheaper than Co. The only drawback of LNO is its structural instability (*i.e.*, structural disorientation and local structure collapse) during charging process (at a highly delithiated state) due to Li/Ni cation-mixing.[10-11] Two main reasons that cause Li/Ni mixing in LNO are: (1) similar ionic radius of $Li^+$ (0.076 nm) and $Ni^{2+}$ (0.069 nm) and (2) partial reduction of $Ni^{3+}$ to $Ni^{2+}$ due to nonstoichiometric structures of LNO.[12] Both these factors combined result in the migration of $Ni^{2+}$ from the transition-metal site towards the Li site during synthesis and delithiation (which also blocks Li diffusion pathway).[13]

Two types of doping methods reported in the literature to reduce cationic disorder are: (a) doping in the metal-oxide block and (b) doping in the intercalation site. Among them, one of the strategies that holds promise is to dope LNO with other metal atoms (especially, transition metal atoms such as Mn, V, and Ti) in the metal-oxide block.[14-17] In addition, a small amount of cobalt doping[18] and aluminum (Al) doping[19] found to improve both thermal instability and electrochemical properties of LNO. In this regard, $LiNi_{0.80}Co_{0.15}Al_{0.05}O_2$ (lithium-nickel-cobalt-aluminum-oxide; NCA) cathode with 15% Co content achieves ubiquitous commercial success due to its high usable discharge capacity and long storage calendar life as compared to LNO.[20] Practical usage of NCA as cathode material is only limited by its rapid capacity fading during charge/discharge cycling,[10-11] which, as mentioned earlier, is due to mixing of Li/Ni cation.[12, 21] Several studies have been reported on the strategies to improve the stability of NCA materials.[16-17, 22-23] These studies are mainly based on various cationic doping in Ni-rich cathode materials to cease the migration of $Ni^{4+}$ ions towards Li-ion site from the transition metal sites. Moreover, doping in the Li-ion intercalation site is also a very important strategy in order to reduce the structural disorientation. Among them, magnesium (Mg)[22-23] and Sodium (Na)[24-27] are the most well-studied dopants which can be embedded into the Li-ion site.

Ceder et al.[28] have demonstrated by ab initio calculations on lithium-nickel-manganese-oxide that the broadening of the *inter*-layer space between consecutive metal oxide layers by doping with a larger ion such as $Na^+$ compared to $Li^+$ helps in reducing the Li/Ni cation mixing and also eases Li-ion diffusion process. This has inspired several studies on doping of various intercalated cathode materials with Na including lithium-vanadium-fluorophosphate, lithium-nickel-cobalt-manganese-oxides, lithium-manganese-oxides, and most importantly in NCA materials.[24, 27] Another reason for choosing Na as a dopant is its abundance on earth's crust that makes it inexpensive.[29-30] Though Na-doped cathodes have slightly lower energy density compared to the un-doped Li-ion cathodes, it improves cycle life,[24, 27] which is a higher priority for large-scale grid applications.[31]

With increased environmental and safety concerns, the recent focus[32] has been on the high energy density and high voltage electrode materials with low toxicity and high thermal stability. At present, the battery industry utilizes 41% of global cobalt supplies, according to the cobalt Development Institute (CDI)[33], and the prediction is that this demand will increase up to 65% within the next decade.[34] On the other hand, cobalt is the side–product of copper (~1.3%) and nickel (~6.7%) mines and paramount portions (~94%) of cobalt supplies are totally dependent on global demand of nickel and copper (Cu).[35] This level of enormous demand for cobalt will affect the market and thus eventually increases its price by a large amount. Hence, reducing the Co content in the (NCA) cathode materials not only addresses the environmental issues to some extent (by decreasing the toxic substances in landfills) but also reduces the cost. NCA materials are one of the promising examples in this direction and are regarded as one of the most appealing materials due to their comparable operating voltage, energy density, and ideal properties for practical applications.[19, 36-37] Tesla motors successfully employed NCA material with 80% Ni and 15 % Co ($LiNi_{0.80}Co_{0.15}Al_{0.05}O_2$).[38-39] Studies on LIBs so far have been mainly focusing on the synthesis and characterization of various cathode materials with high operating voltage window and high energy density.[40-42] However, according to the best of our knowledge, very few reports exist on the effect of replacing Co with Ni on the structural stability and the electrochemical properties of cathode materials. For example, one of the recent study reports about Ni-rich NCM cathode materials.[43]

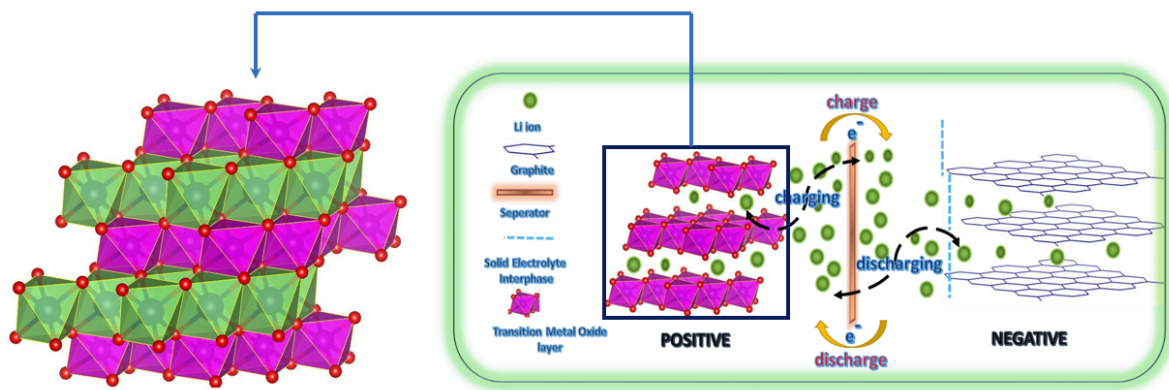

**Figure 1.** Representation of charge-discharge mechanism in lithium-ion battery. Typical layered crystal structure of cathode material; LMO (LCO/ LNO) [inset, zoomed in left]. Here the green signifies the Li/Na site and purple signifies transition metal site (Co/Ni).

Thus, our primary focus is to predict an NCA composition with lowest possible percentage of Co without any significant loss of electrochemical performance and structural stability. Our objective is to understand the structural changes and electrochemical behavior of NCA materials as a function of Co concentration using first principle approach. To this end, four different combinations were considered by decreasing Co concentration ($z$) and replacing it with Ni atoms ($y$). These four different materials are in between the range of 16.66% to 4.16% of Co content. Electrochemical properties, such as intercalation potential[44] (V vs. Li/Li$^+$) as a function of theoretical capacities[45] for each of the four materials ($z_1 = $ **16.66%**, $z_2 = $ **12.5%**, $z_3 = $ **8.33% and** $z_4 = $ **4.16%**) were studied along with the impact of the same on the overall structure. Finally, the influence of Na doping in Li-site on two of the materials, with highest and lowest Co content was also investigated. Some guidelines for designing stable, cost-effective, and environmentally benign cathode materials in LIBs were provided.

## 2. Computational Methodology

All the electronic structure calculations were performed using the Density Functional Theory (DFT) method as implemented in Vienna Ab initio Simulation Package (VASP).[46] Projector augmented wave (PAW) pseudopotential is taken for the inert core electrons, and valence electrons are represented by plane-wave basis set.[47-48] The generalized gradient approximation (GGA), with the Perdew–Burke–Ernzerhof (PBE)[49] exchange–correlation functionals, is taken into account. We have performed DFT+U[8] calculations using the generalized gradient approximation (GGA), with the Perdew–Burke–Ernzerhof (PBE)[9] exchange–correlation functionals. The U parameters for Ni and Co were chosen from the previous studies, and their values were 6.2 eV[50] and 3.32 eV[51] for Ni and Co respectively. The plane wave basis was set up with a kinetic energy cutoff of 520 eV. Brillouin zone is sampled using gamma-centered Monkhorst−Pack scheme with 2 × 2 × 1 $k$-point grid for each of the NCA and Na-NCA species. All the internal coordinates are relaxed until the Hellmann−Feynman forces are less than 0.02 eV/Å.

We have performed geometry optimization (*i.e.*, relaxing the geometry to obtain the structure with the lowest energy) of each structure with varying Li numbers (***n***). The energy minimization calculation will provide us the information about the difference in Gibbs free energy ($\Delta G$) and lithiation potential $V(n)$. The lithiation potential is defined as[52-54]

$$V(n) = \frac{-\Delta G}{Z_e F} \tag{1}$$

where ***F*** is the Faraday constant, and $z_e$ is the charge (in electrons) transported by Li in the electrolyte. In most non-electronically conducting electrolytes, $z_e = \mathbf{1}$ for Li intercalation. The change in Gibb's free energy is given by the formula[54]

$$\Delta G = G_{Li_{n1}Y} - (G_{Li_{n2}Y} + \Delta n G_{Li}) \tag{2}$$

Where, $\Delta n = n2 - n1$ and it signifies the total number of Li atoms transferred during lithiation/de-lithiation process. Here, $Li_{n_i}$ represents the system with $n_i$ no of Li atoms and $Y$ represents the mixed oxide. In this present study, $n_i$ varies from 0 – 24. $G_{Li}$ is the total Gibbs free energy of a single Li atom in elemental body-centered cubic Li. If the energies are expressed in electron volts, the intercalation potential of the $Li_{n_i}Y$ (V vs. Li/Li$^+$) as a function of Li content can be shown as[44]

$$V(n) = \frac{-\Delta G}{\Delta n} \qquad (3)$$

The composition range over which Li can be reversibly intercalated determines the battery capacity. Theoretical capacity of electrode materials depends on the molar weight and the number of reactive electrons of the specific material. Theoretical capacity (with mAh/g units) has the following formula[45]:

$$C_t = \frac{n_i \times F}{3.6 \times M} \qquad (4)$$

where $n_i$ is the number of reactive electrons in the formula and it is equivalent to the number of Li atoms present in the corresponding computational cell. $M$ represents the molar weight of the active electrode material.

Initially, layered LiNiO$_2$ crystal structure was considered from the available literature[50], and then Al atom was introduced in the crystal structure followed by the Co atoms. The Al and Co doping were carried out via simultaneous incorporation of Al and Co and removal of Ni atoms respectively. In our present study, Li$_n$Ni$_m$Co$_p$Al$_1$O$_{48}$ is taken as a template to represent the following systems: NCA_I (NCA$_{Co=16.66\%}$): Li$_{24}$Ni$_{19}$Co$_4$Al$_1$O$_{48}$; NCA_II (NCA$_{Co=12.5\%}$): Li$_{24}$Ni$_{20}$Co$_3$Al$_1$O$_{48}$; NCA_III (NCA$_{Cso=8.33\%}$): Li$_{24}$Ni$_{21}$Co$_2$Al$_1$O$_{48}$; and NCA_IV (NCA$_{Co=4.16\%}$): Li$_{24}$Ni$_{22}$Co$_1$Al$_1$O$_{48}$ In order to investigate the effect of Na doping at Li-site, one Na atom was incorporated in each of the layered Li-site of NCA_I and NCA_IV. As the considered structure has two Li-layers, overall Na incorporation results in 2 Na atoms doping in each structure. 2 Li atoms, each from two different layers were replaced by 2 Na atoms to generate Na_NCAs. In case of Na_NCA (Li$_{n-2}$Na$_2$Ni$_m$Co$_p$Al$_1$O$_{48}$), we have only considered two structures with highest and lowest Co content and these are the following materials: Na_NCA_I (Na_NCA$_{Co=16.66\%}$): Li$_{22}$Na$_2$Ni$_{19}$Co$_4$Al$_1$O$_{48}$; and Na_NCA_IV (Na_NCA$_{Co=4.16\%}$): Li$_{22}$Na$_2$Ni$_{22}$Co$_1$Al$_1$O$_{48}$. In order to account for the structural deformation of these systems, AIMD (Ab Initio Molecular Dynamics) simulation was performed under a canonical NVT ensemble at room temperature (RT) for ~5000 fs. The temperature of 298 K (RT) is maintained by a Nose Hoover thermostat.

## 3. Results and Discussion

Intercalation/insertion property is the basic requirement for a solid-state material towards its usefulness as a cathode material. The easy intercalation/deintercalation of cations (Li$^+$ ions) without causing any non-reversible structural changes in the host material is responsible for its success and widespread use.[9]

**3.1 Effect of Co content on the electrode potentials and structural stability of Li$_n$Ni$_m$Co$_p$Al$_1$O$_{48}$ Materials**

Optimized structures of fully lithiated NCA_I through NCA_IV (*i.e.*, Li$_{24}$Ni$_{19}$Co$_4$Al$_1$O$_{48}$, Li$_{24}$Ni$_{20}$Co$_3$Al$_1$O$_{48}$, Li$_{24}$Ni$_{21}$Co$_2$Al$_1$O$_{48}$ and Li$_{24}$Ni$_{22}$Co$_1$Al$_1$O$_{48}$), are shown in **Figure 2a**. The decrease in Co concentration along with the simultaneous increment in Ni concentration happens from NCA_I to NCA_IV while keeping the Al concentration constant. In this way, it will be easier to spot characteristic differences between these NCAs due to change in Co concentration. The most basic characteristic of any cathode material is its ability to enable Li intercalation process while maintaining the stability of the metal oxide framework. Some materials with high intercalation potential are not suitable for practical purpose due to their structural instability. In order to account for the electrochemical performance via computational studies, the changes in intercalation potential values were calculated for the different state of charge for four NCA materials with different Co concentrations. Starting from the active cathode materials (fully lithiated NCAs; Li = 24 in a fully lithiated state for all of the four cases), 4 Li atoms were removed at a time (~16.66% Li) in order to generate the consecutive delithiated structures. Each of these structures was fully relaxed along with the cell relaxation in order to study any structural deformation during the stepwise delithiation process. **Figure 2b** represents the fully delithiated (when the number of Li = 0) optimized structures for all of the four NCA materials. The theoretical specific capacities for each NCAs vary around ~ 278 mAh/g, in which NCA_IV has the highest value (278.249 mAh/g), while NCA_I has the lowest value (278.163 mAh/g) due to a slightly higher molecular weight of Co than Ni. The typical specific capacity for NCA (~279 mAh/g) is in good agreement with our calculated values,[55] but, on the other hand, the practically achievable specific capacity is almost around ~200 mAh/g.[55] It indicates that complete delithiation does not take place during the practical experimental setup. **Figure 3** represents the intercalation potential curve as a function of theoretical specific capacity. The intercalation potential for each of the NCA materials is in good agreement with the available literature value for the most well-studied NCA ($LiNi_{0.80}Co_{0.15}Al_{0.05}O_2$).[56] From **Figure 3**, it is clear that all of the NCA materials are following the similar nature as the discharge potential curve for the available cathode materials.[20, 56] The intercalation potential for each case shows a steady decrease with the increment in the specific capacity. The most important observation is that for each of the NCAs, the average potential values do not differ much and it ranges from 4.36 V to 3.72 V. Among them, NCA_I (with highest Co content in this present study) shows the highest average potential and the lowest value is for the NCA_IV. As most of these cathode materials did not undergo delithiation beyond 50% due to potential risk of structural instability, it is essential to note the potential at 50% of Li content (when Li=12). In our study, the intercalation potential decreases with a decrease in Co content. At 50% delithiated state, the values for all the NCAs are 4.63 V (NCA_I), 4.17 V (NCA_II), 4.11 V (NCA_III), and 3.84 V (NCA_IV). At 50% of Li content, around 17 % drop in intercalation potential was observed going from NCA_I to NCA_IV. On the other hand, the average potential drop is only ~14.7%.

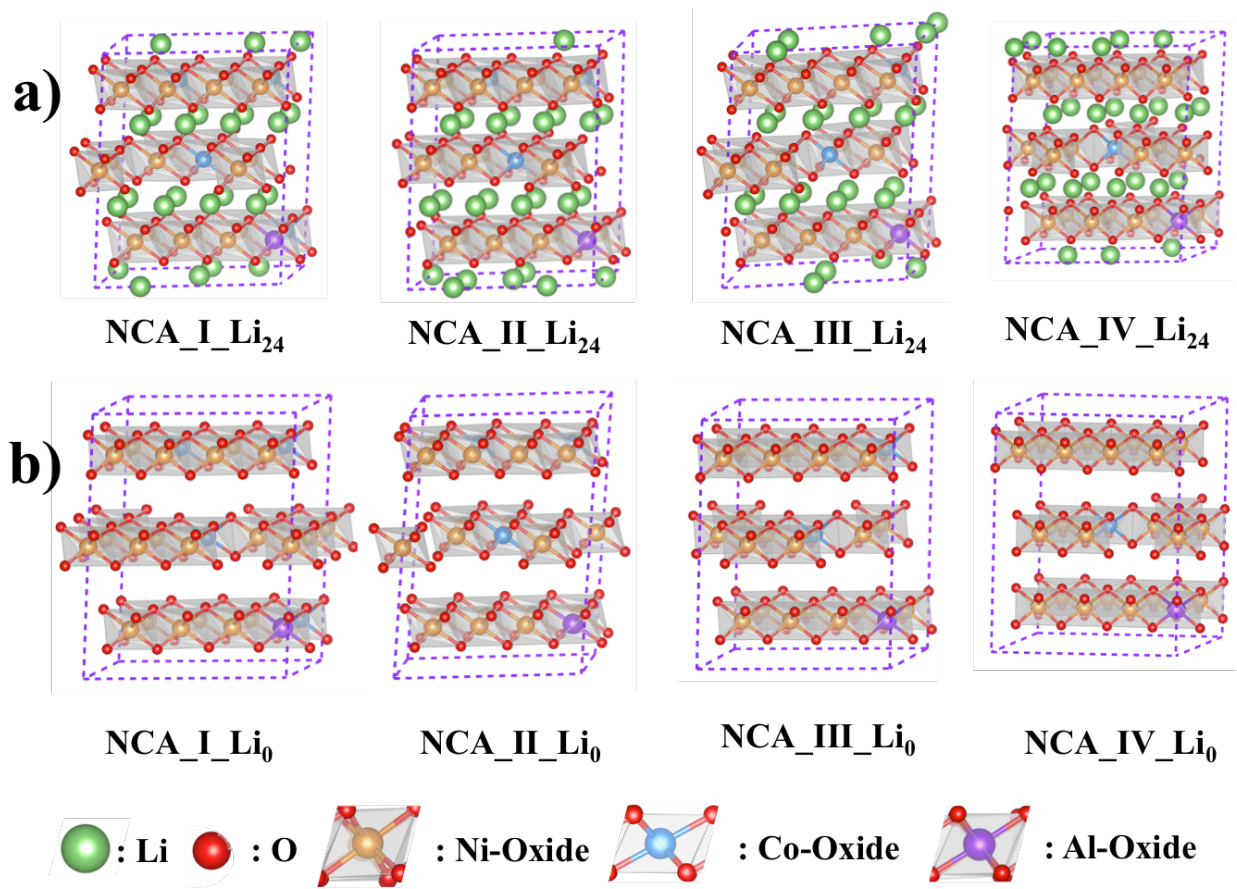

**Figure 2.** Pictorial representation of the optimized structures of a) fully lithiated NCA_I, NCA_II, NCA_III and NCA_IV; and b) completely delithiated NCA_I, NCA_II, NCA_III and NCA_IV

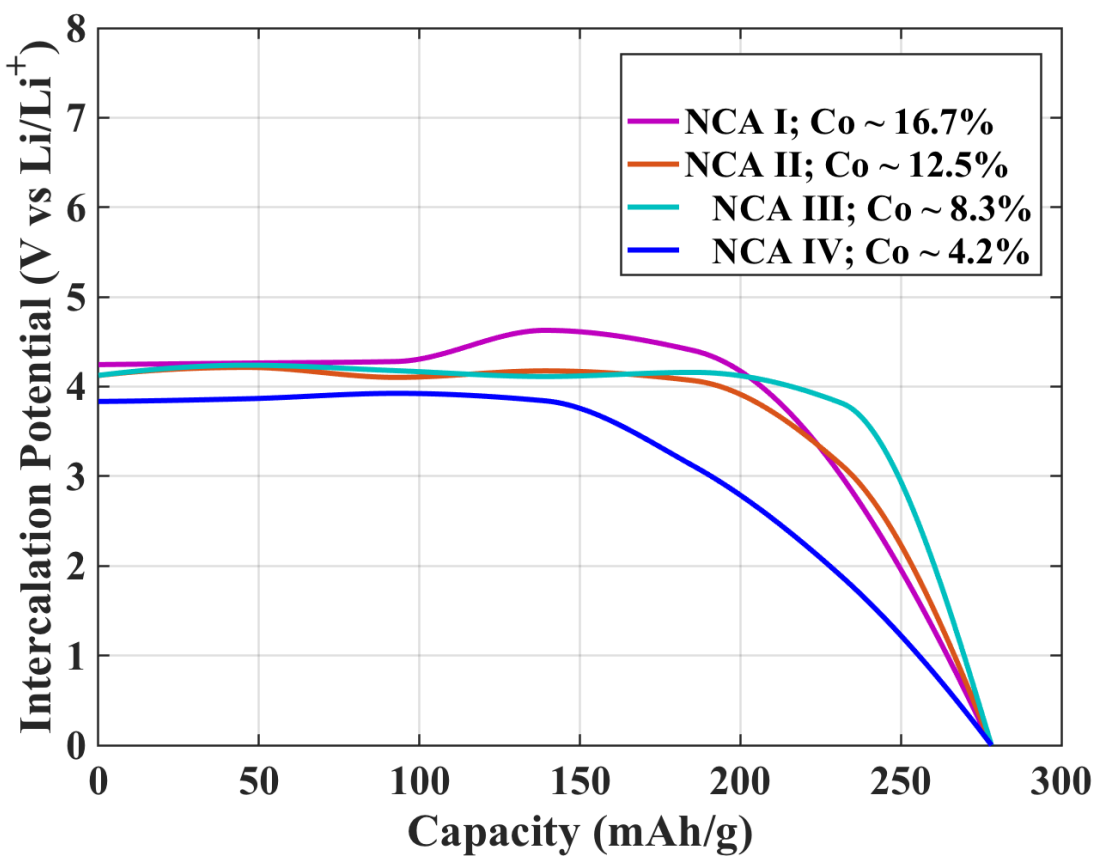

**Figure 3.** Comparative plot of intercalation potential (V vs. Li/Li$^+$) with varying theoretical specific capacity for NCA_I, NCA_II, NCA_III and NCA_IV

With the increase in Ni concentration, ion-mixing phenomenon described in the introduction, start playing its role to affect the structural stability. Increase in Ni concentration is known to affect the structural stability of NCA materials at highly delithiated states[57-58]. In order to study the structural changes due to decreasing Co content, we have considered the three extreme cases for NCA_I and NCA_IV and compared their internal bond distances: (i) fully lithiated state, where ***n* = 24**; (ii) 50% delithiated, where ***n* = 12** and (iii) completely delithiated state, where ***n* = 0**. At the beginning, we started our analysis by computing internal bond distances (see **Table S1** of the supporting information) and volume changes during the delithiation process (see **Table S3** of the supporting information). We have tabulated the changes in average *intra*-layer bond distances of Ni–O, Co–O and Al–O for NCA_I and NCA_IV with decreasing Li–content. Note, from the **Table S1,** that the average metal–oxygen bond distances decrease steadily with a decrease in Li–content, which is true for both NCAs; this decrease can be attributed to a stronger covalent nature of the metal–oxygen in the absence of Li. A steady decrease in the average non–bonded *intra*–layer Ni–Ni distances are also observed. On the other hand, the average non–bonded *inter*–layer distances are found to show increment with decreasing Li–content. The strength of Li–O interaction decreases with the decrease in Li content resulting in the predictable increase of the *inter*-layer distances between metal-oxide layers. From our calculated bonded and non–bonded distances, it is pretty clear that with the decrease in Li–content, the average *intra*–layer distances decrease, but the average *inter*–layer distances increase. The lengthening of *inter*-layer distances of delithiated NCAs is evident from the fact that there is no Li left to form the bond with the oxygen and thus the layers are not attached through chemical bonds. Thus, the absence of Li gives oxygen the liberty to form stronger bonds with the other metals present in the oxide layers resulting in the shortening of M–O (where M = Ni/Co/Al) bonds in case of NCA_Li$_0$ as compared to the NCA_Li$_{24}$. In case of NCA_I and NCA_IV, the volume changes (see **Table S3** of the of the supporting information) for complete delithiation are ~0.32% and ~2.78% respectively. However, the volume changes at the stage of 50% delithiation are only 0.34% and 0.39% for NCA_I and NCA_IV respectively. Further, we have performed room temperature AIMD simulations of the NCA_IV_Li$_{0/12/24}$ (i.e. three cases with number of Li atoms = 0, 12 and 24) systems for ~5100 fs. Our simulation did not account for any significant structural disorientation even after complete delithiation (see **Figure 4**). Although, we have observed small changes in average *inter*-layer distances. However, it is evident that oxide layers in NCA_IV is not falling apart even in the fully delithiated state. Therefore, we should not dismiss the possibility of its usage as a cathode material. Thus, our DFT study demonstrates the scope of modification of NCA materials via systematic study of the effects of stepwise reduction of Co content. However, various other possibilities are still open to research for further modification. One such possibility is discussed in the next section.

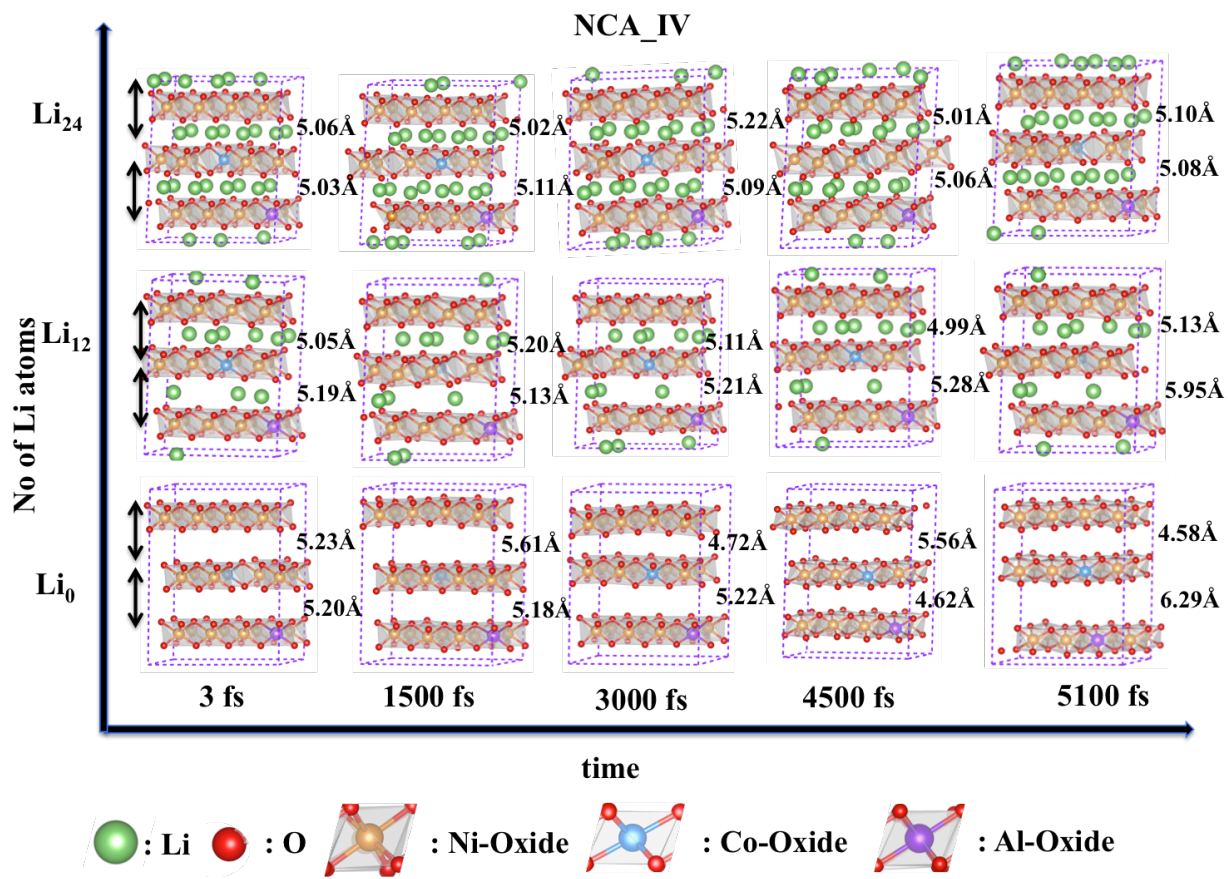

**Figure 4.** Comparative room temperature AIMD simulation images of NCA_IV_Li$_0$, NCA_IV_Li$_{12}$ and NCA_IV_Li$_{24}$ at different time frames.

## 3.2 Effect of Co content on the electrode potentials and structural stability of Li$_{n-2}$Na$_2$Ni$_m$Co$_p$Al$_1$O$_{48}$ Materials

As mentioned earlier,[24, 27] a small amount of Na doping in the Li site has been recognized to resist structural deformation during delithiation (e.g., interlayer collapse, Li/Ni mixing), but it has lower energy density compared to the un-doped Li-ion cathode. In our study, we have incorporated a very small amount of Na (2 Na atoms per formula unit) in two extreme cases, such as NCA_I and NCA_IV in order to study its effect in NCA materials with decreasing Co content. To study the electrochemical properties, we have considered Na-doped NCA materials (fully lithiated active materials) with following formula units: (1) Na_NCA_I: Li$_{22}$Na$_2$Ni$_{19}$Co$_4$Al$_1$O$_{48}$; and (2) Na_NCA_IV: Li$_{22}$Na$_2$Ni$_{22}$Co$_1$Al$_1$O$_{48}$. It is essential to mention here that successful synthesis of similar compounds is recently reported by Hu et al.[27] Optimized structures of Na_NCA_I and Na_NCA_IV in fully lithiated and completely delithiated forms are shown in **Figure 5a** and **Figure 5b** respectively. The theoretical specific capacities for both the Na_NCAs vary around ~251 mAh/g, in which Na_NCA_IV has the highest value (251.569 mAh/g) and Na_NCA_I is the one with the lowest value (251.492 mAh/g) due to a slightly higher molecular weight of Co than Ni. The decrease in specific capacity is due to the presence of heavier Na atoms and presence of fewer numbers of Li atoms per formula unit. Hence, the overall theoretical capacities are lowered by a minimal amount upon Na doping (~9%).

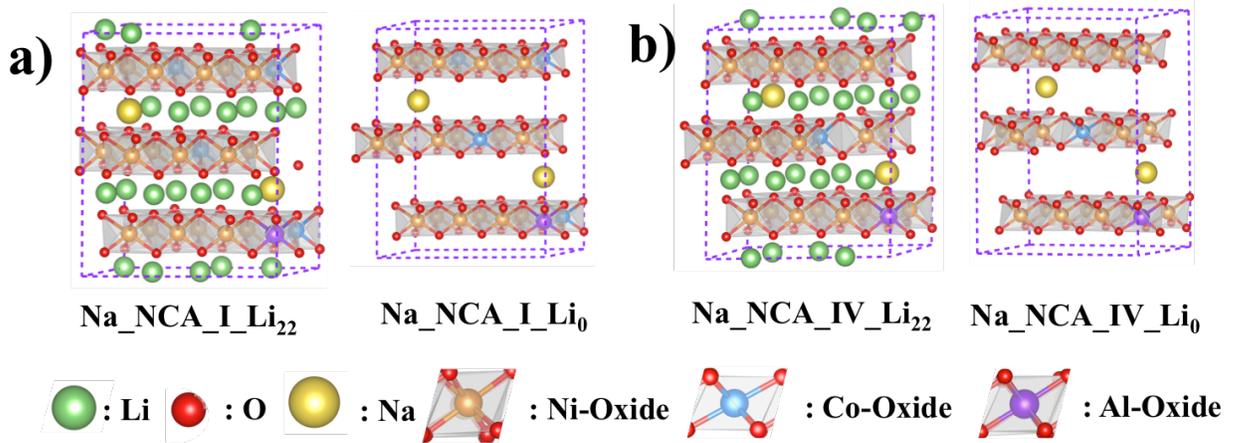

**Figure 5.** Pictorial representation of the optimized structures of a) fully lithiated Na_NCA_I, and Na_NCA_IV; and b) completely delithiated Na_NCA_I, and Na_NCA_IV.

**Figure 6** represents a comparative curve of intercalation potential (V vs. Li metal potential) as a function of specific capacity for Na_NCAs. From **Figure 6**, it is clear that each of the Na_NCA materials is following the similar nature as the discharge potential curve for the available cathode materials. The intercalation potential for each case shows a steady decrease with the increment in the specific capacity as similar to their NCA counterparts (**Figure 3**). The average potential values observed for Na_NCA_I and Na_NCA_IV are 4.68 V and 4.14 V respectively and apparently, there is only 11.5% drop in the overall average potential value by going from Na_NCA_I to Na_NCA_IV. Moreover, an overall increment in intercalation potential is also noticeable for the Na counterparts as compared to their non-Na counterparts. For the cathode materials with the lowest Co content (p = 1; NCA_IV and Na_NCA_IV), Na_NCA_IV shows an increment of ~11.3% of the overall average potential as compared to the NCA_IV. This observation is expected from the knowledge gained regarding the presence of bigger Na in Li-site, which eventually eases up Li intercalation. So, in spite of a slight decrease in the theoretical specific capacity, the Na-doped NCAs are one of the viable options from the point of view of increased intercalation potential. Moreover, Na doping in Li site has been distinguished as a beneficial strategy to reduce the electrochemical polarization effect resulting in stabilization of the structure as compared to the un-doped species.[24, 27] Na-doped structures are also identified to possess enhanced surface properties and improved cycle performance.[24]

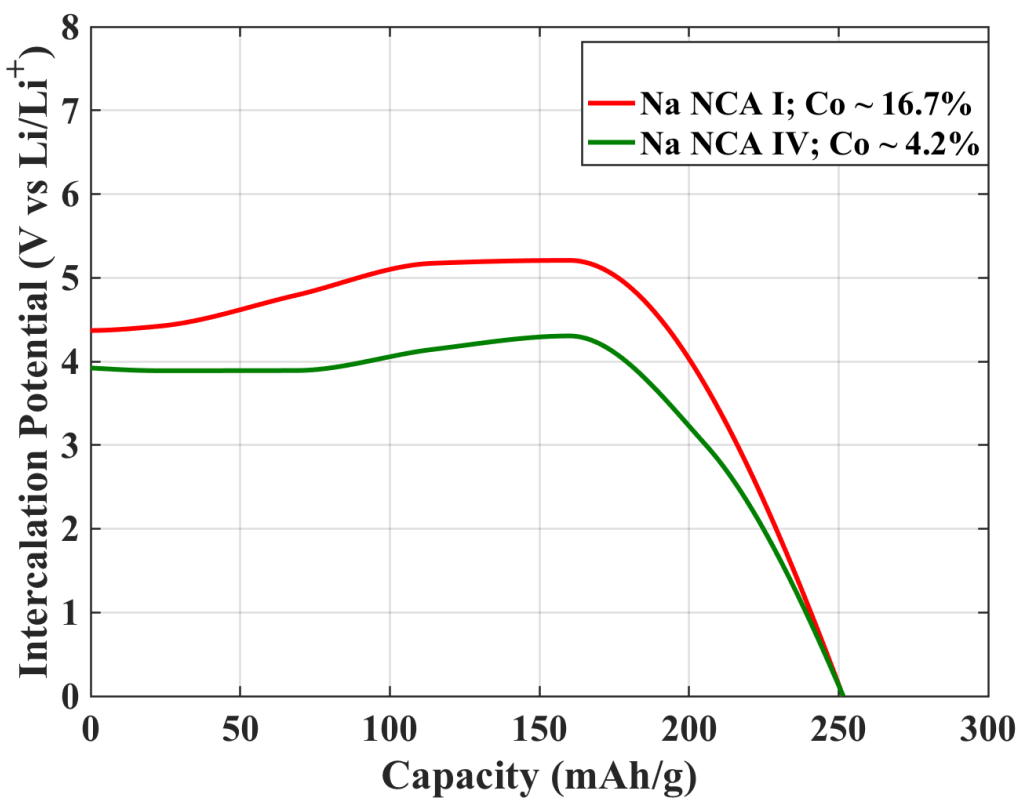

**Figure 6.** Comparative plot of intercalation potential (V vs. Li/Li$^+$) with varying theoretical specific capacity for Na_NCA_I and Na_NCA_IV.

We have also compared the various structural features of the Na_NCA_I and Na_NCA_IV in order to study the effect of decreasing Co content on the Na-doped materials. *Intra*-layer and *inter*-layer average bond distances of the fully lithiated, 50% lithiated and completely delithiated Na_NCAs are tabulated in **Table S2** of the supporting information file. The *inter*-layer Ni–O, Co–O and Al–O bond distances in both the Na_NCAs show a steady decrease upon delithiation. The similar trend is followed by NCAs as well (see **Table S1** of the supporting information). It is obvious from the fact the Metal (M) and Oxygen forms stronger bonds in the absence of Li due to the lack of Li–O bond. However, in case of the Na_NCAs, the M–O bonds are not getting shorter upon delithiation as in the case of NCA. This is probably due to the presence of existing Na–O bond even after the complete delithiation. The same is true for Ni–Ni *intra*-layer bond distances as well. The most important observation is the extent of increment in the average *inter*-layer Ni–Ni bond distances (see **Table S1** and **Table S2** of the supporting information). The increment of the average *inter*-layer Ni–Ni bond distances is ~ 11.5 % going from NCA_IV to Na_NCA_IV. This is due to the presence of Na in Li-site even after complete delithiation. Moreover, the presence of large Na atoms are known to increase the volume of the host materials in large extent and so happens in our case as well (see **Table S3** of the supporting information).[59] It is important to mention here that the changes in volume upon delithiation in the present Na_NCA systems are more than NCA systems. However, the magnitude of volume changes for Na_NCAs is not very high and they are only ~4.7% and 6.08% for Na_NCA_I and Na_NCA_IV respectively. Further, room temperature AIMD simulations of the Na_NCA_IV_Li$_{0/12/24}$ (i.e. three cases with number of Li atoms = 0, 12 and 24; see Figure 7 a-c respectively) systems for ~5100 fs did not suggest any significant structural changes even after the complete delithiation. Although, the average *inter*-layer distances underwent changes during delithiation and throughout the AIMD simulation, the overall structure is not collapsing. The *inter*-layer distances in Na_NCAs are observed to be more as compared to the NCAs and it is due to the presence of Na atoms in the Li-site even after the complete delithiation. The most important observation is that at the end of our AIMD simulation for Na_NCA_IV, we found that distances between the metal oxide layers are far more uniform than that of NCA_IV (see **Figure 4** and **7**). The uniformity in the structure is due to the presence of 1 Na atom in each Li layer indicating the existence of Na–O bonding. It eventually prevents the oxide layers from falling apart even in a highly delithiated scenario. The reason behind the lower *inter*-layer distances in case of Na_NCA_IV_Li$_{22/10}$ (i.e. two cases with number of Li atoms = 22 and 10) is due to the presence of stronger Li–O bond along with the weaker Na–O bond. The *inter*-layer distances increase in the case of Na_NCA_IV_Li$_0$ (number of Li atoms = 0). Though Na helps to keep the uniformity in the *inter*-layer distances, the increment in distances takes place due to the weaker nature of Na–O bond as compared to the Li–O bond.[60] It is obvious from our study that NCA_IV lacks this stability gain from Na. Thus, our calculations explore the possibility of simultaneously introducing two major modifications in NCA materials, which are Na doping in the intercalation site and effects of Co reduction.

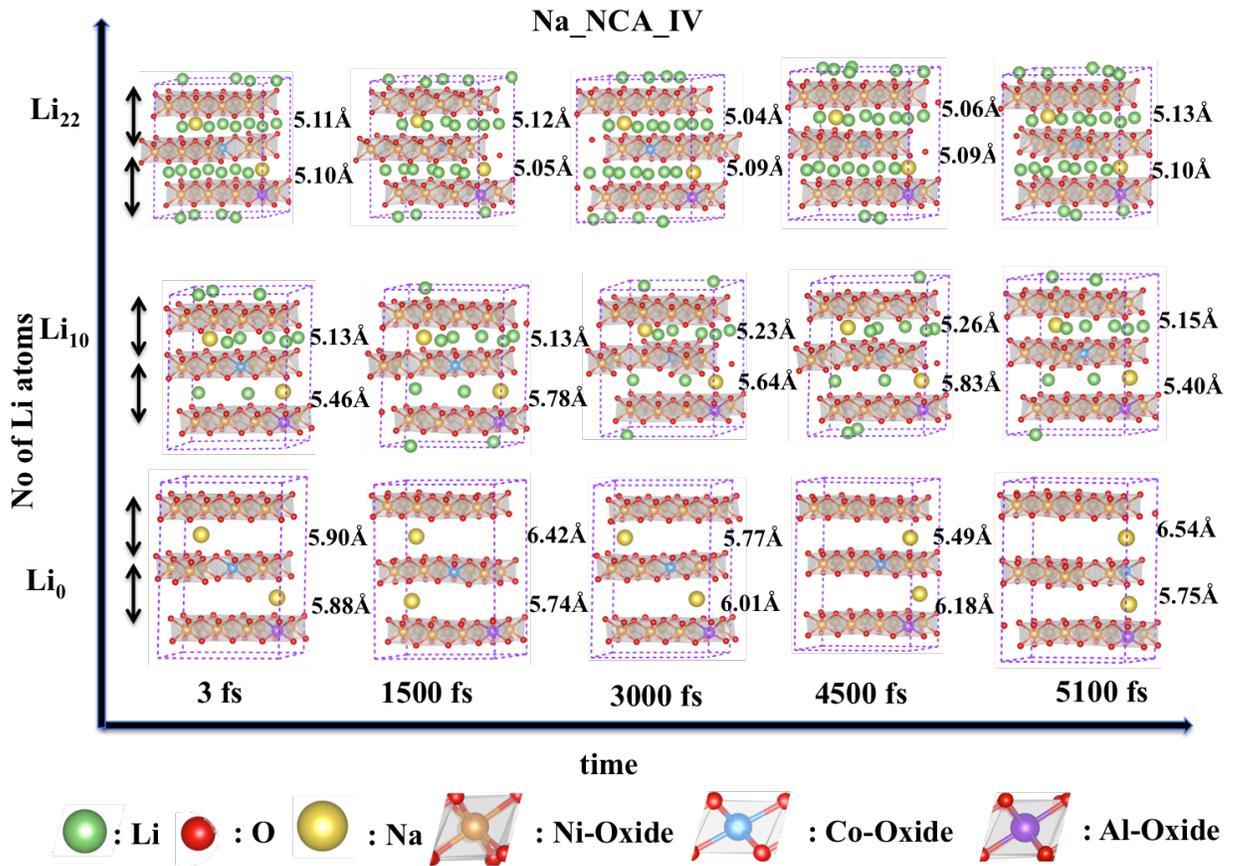

**Figure 7**. Comparative room temperature AIMD simulation images of NCA_IV_Li$_0$, NCA_IV_Li$_{12}$ and NCA_IV_Li$_{24}$ at different time frames.

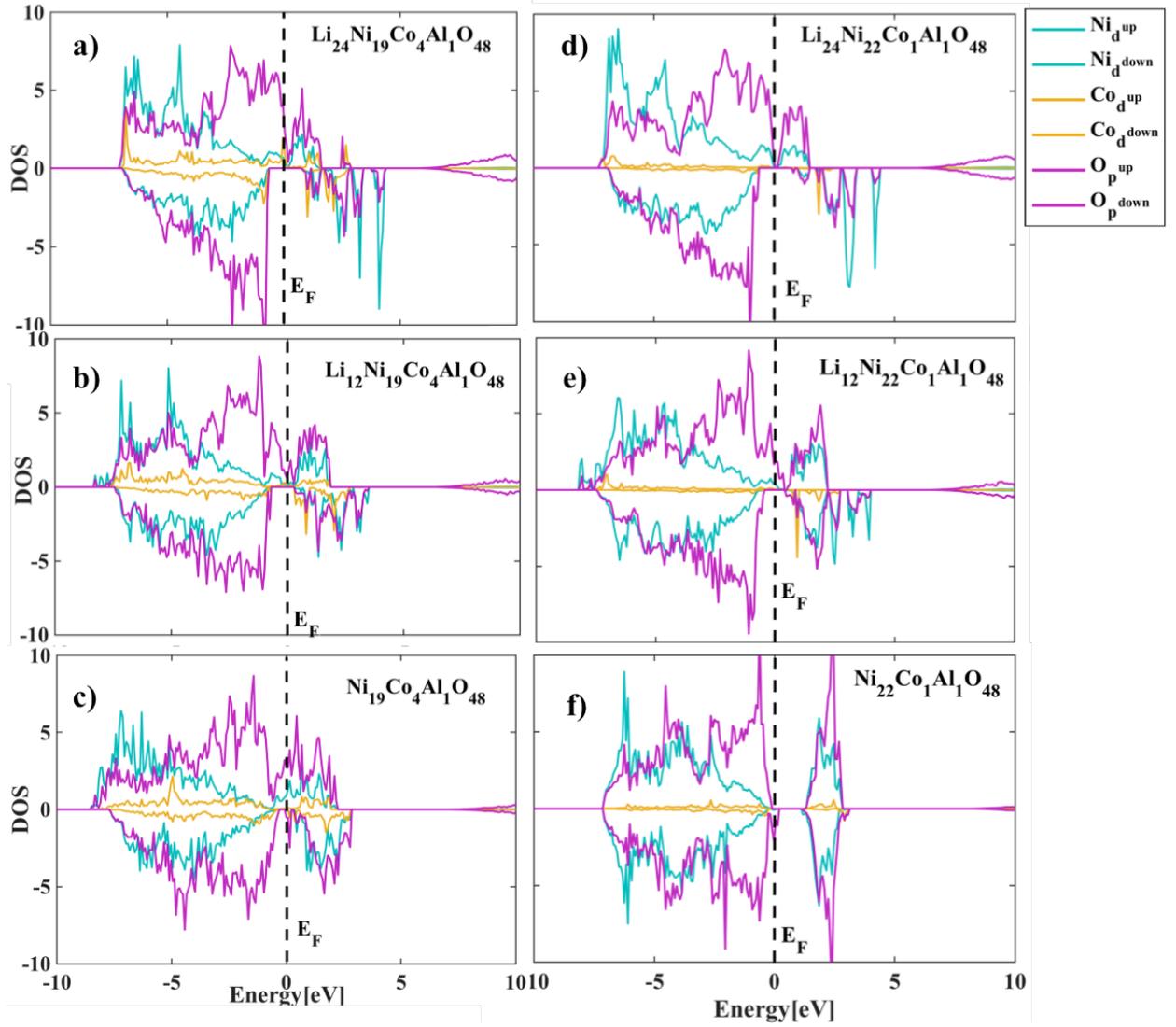

**Figure 8**. Schematic projected density of states (PDOS) for i) NCA_I (from **a** to **c**) and (ii) NCA_IV (from **d** to **f**).

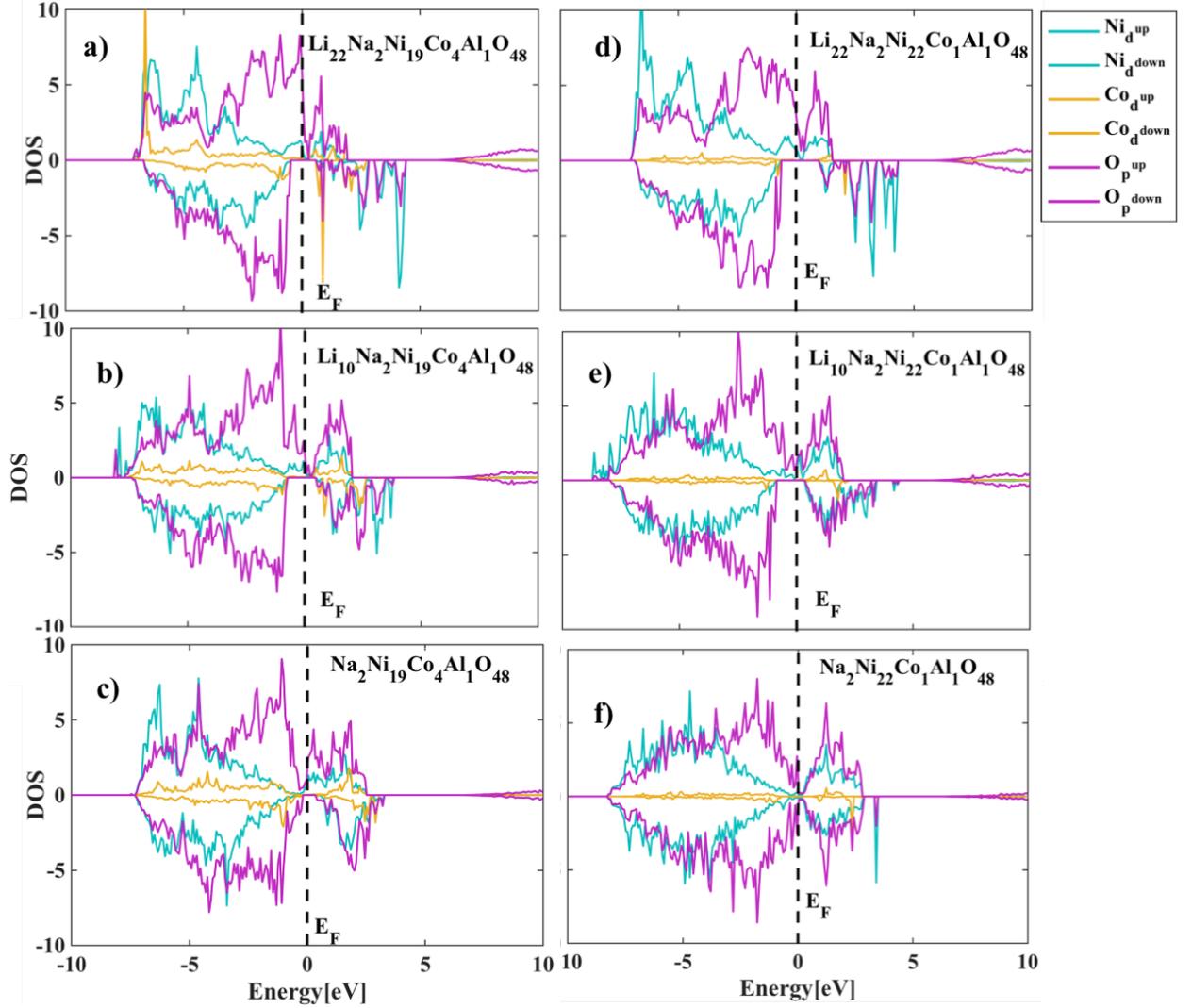

**Figure 9**. Schematic projected density of states (PDOS) for i) Na_NCA_I (from **a** to **c**) and (ii) Na_NCA_IV (from **d** to **f**).

### 3.3 DOS analysis of $Li_nNi_mCo_pAl_1O_{48}$ and $Li_{n-2}Na_2Ni_mCo_pAl_1O_{48}$ Materials

In order to understand changes in electronic properties during the course of lithiation/delithiation, we also have incorporated Density of States (DOS) analysis for NCAs and Na_NCAs (see **Figure 8** and **9** respectively). In this present study, we have considered NCA_I, NCA_IV, Na_NCA_I and Na_NCA_IV at 100% ($Li_{24/22}$) (i.e. number of Li atoms = 24 for 100% lithiated NCA_I/IV and number of Li atoms = 22 for 100% lithiated Na_NCA_I/IV), 50% ($Li_{12/10}$) (i.e. number of Li atoms = 12 for 50% lithiated NCA_I/IV and number of Li atoms = 10 for 50% lithiated Na_NCA_I/IV) and 0% ((Li$_0$) (i.e. number of Li atoms = 0 for fully delithiated NCA_I/IV and Na_NCA_I/IV) Li concentrations. The atomic DOS distributions will give us an idea about the combined effects of the decrease in Co content and the simultaneous increase in Ni content along with the overall reduction of Li content on electronic properties, in case of NCAs and Na_NCAs. In case of NCA_IV, the most important observation is that at 100% delithiated state (Li$_0$), the behavior of the local DOS around Fermi level is getting changed and indicating an insulator kind of behavior (**Figure 8f**). At 50% delithiated state (Li$_{12}$), NCA_IV shows metallic behavior (**Figure 8e**) and it is well known fact that it is hard to delithiate cathode materials beyond 50% from the practical purpose due to possible structural deformation.[61] So, NCA_IV also remains as a viable option as a suitable cathode material. NCA_I maintains the metallicity at each stages of delithiation (**Figure 8a-c**). In this case, due to the presence of more numbers of cobalt atoms, a significant overlap between oxygen's 2$p$ orbital and cobalt's 3$d$ orbital takes place near Fermi level and thus makes the electron density continuous and eventually maintains the metallic behavior throughout. Moreover, the higher Co content shifts the Fermi level (drags electron density closer to t$_{2g}$ orbitals from e$_g$) of mixed transition metals closer to the oxygen's 2$p$ orbitals and thus accounts for the higher potential.

The electronic structure of cathode materials has a substantial impact on the voltage. In LiMO$_2$ (where MO$_2$ represents mixed metal oxide), the average oxidation state of the transition metals (Co and Ni in this case) is 3+.[62] In the fully lithiated state, Co$^{3+}$ (3$d^6$) acquires a low-spin configuration (t$_{2g}^6$e$_g^0$) in an O$^h$ environment of oxide ions.[9] The t$_{2g}$ orbitals are energetically close to the oxygen 2$p$ orbitals and thus signify stronger overlap between these orbitals. On the other hand, Ni$^{3+}$ (3$d^7$) adopts a low-spin (3$d^7$) configuration (t$_{2g}^6$eg$^1$)[9], but full occupancy of the t$_{2g}$ levels ensures that one must focus on the single electron of the e$_g$ orbitals. Here overlap between e$_g$ orbitals with oxygen's 2$p$ orbitals also gives rise to strong interactions. In case of higher Ni content, Fermi level resides near the higher energy e$_g$ orbitals, whereas, Fermi level resides near the lower energy t$_{2g}$ level in case of systems with higher Co content. These locations of Fermi level results in lower intercalation potential for materials with high Ni content (high Fermi level) as compared to the materials with higher Co content. Previous studies also agree with this explanation.[2, 63-66] Extraction of Li atoms are known to be associated with the partial oxidation phenomenon of $Ni^{3+}$ and $Co^{3+}$.[9] Partial oxidation of $Ni^{3+}$ ($3d^7$ configuration; $t_{2g}^6 e_g^1$) to $Ni^{4+}$ ($3d^6$ configuration; $t_{2g}^6 e_g^0$) decides the position of Fermi level, which is near $t_{2g}$ orbitals due to the vacant $e_g$ orbitals. In case of $Co^{3+}$ ($3d^6$ configuration; $t_{2g}^6 e_g^0$), partial oxidation to $Co^{4+}$ ($3d^5$ configuration; $t_{2g}^5 e_g^0$) maintains the position of Fermi level close to oxygen 2p band.

Most importantly, DOS images of Na_NCA_IV show an overall continuation of the electron density near the Fermi level even at the completely delithiated state (**Figure 9f**). This observation can be explained from the fact that during delithiation, the partial oxidation of $Ni^{3+}$ and $Co^{3+}$ reduces in some extent due to the presence of Na atoms in the Li-site. Thus Na_NCA_I and Na_NCA_IV show almost equivalent DOS images over the entire course of lithiation/delithiation. The lesser the extent of overall changes, such as structural, electronic and intercalation potential, more suitable will be the materials to use for the practical purpose. In this order, Na_NCAs, specially Na_NCA_IV with very low Co content is one of the lucrative alternatives. Low cost of Na due to its abundance compared to Li[67] and prediction of the scarcity of Li resources in the foreseeable future are other essential aspects of inserting Na in the rechargeable Li-ion batteries for large-scale applications.[68] Thus, despite its low energy density, Na ion is known as an adequate alternative due to its abundance, low-cost, safety, and environmental non-lability.[69-71] In this study, we have incorporated a very small fraction of Na in order to account for the changes with respect to the electrochemical properties of our studied NCA materials. But in future, a higher fraction of Na incorporation can also be beneficial to reduce the fraction of Li for large-scale applications. The decrease in Co concentration from Na_NCA_I/NCA_I to Na_NCA_IV/NCA_IV is ~75%, whereas, increase in Ni concentration for the same is only 15.8%. Overall, this reduction in Co concentration significantly reduces the total cost by ~28.7 USD/lb. Detail discussion regarding cost reduction is provided in the **section I** of the supporting information. Our results envisage an appealing strategy of modification of the NCA materials from the point of view of large-scale applications.

Our study demonstrates a strategic approach in order to design a modified NCA material via step by step reduction of Co concentration and utilize the added advantage of Na doping. Our sole objective is to provide a clear understanding of the overall phenomenon from the point of view of DFT methodology. It is important to mention that we have not considered all the phenomena[72] occurring in the electrochemical cell and thus our calculated potential values differ slightly from the experimental values. To be specific, we have not considered the probable effect of electrolyte used in the particular cell. Moreover, oxidation and delithiation of cathode materials occur simultaneously with the reduction and lithiation of anode materials (Graphite). Thus the overall intercalation potential depends on both the cathode and anode materials, and we have not considered any effect concerning the anode material.[8]

## 4. Conclusion

In summary, we have performed DFT calculations to investigate four types of NCA cathode materials with generalized formula **$Li_nNi_mCo_pAl_1O_{48}$.** In addition, we have also investigated two types of sodium doped NCAs, i.e., Na_NCAs with the generalized formula **$Li_{(n-2)}Na_2Ni_mCo_pAl_1O_{48}$.** We have compared our calculated potential and established that the trend is similar to the previously published literature. The decrease in Co content (from NCA_I to NCA_IV) marginally reduces the overall intercalation potential for each value of Li concentration, but this extent of decrease becomes negligible in case of practical and economical purposes. For example, the overall potential drop is only ~14.77% while the change in specific capacity is insignificant (in the range of ~278 mAh/g for all four NCA structures). However, the decrease in cobalt concentration connotes a considerable cost reduction for the battery industries and shrinks the consequences of cobalt–poisoning accordingly. We have also explored the possible structural deformation (bonding and non-bonding distances of the layered material) of NCA_IV in case of 100% delithiation and incorporated the comparative bond distances (*inter* and *intra*-layer) of fully lithiated and fully delithiated moieties in our study. It can be quantified from our calculations that overall changes in the bond distances are inconsequential. Finally, the role of sodium doping in the NCA type cathode materials is extensively investigated due to its role to decrease the Li/Ni cation mixing during delithiation. Moreover, the decrease in overall potential value for Na_NCAs with decreasing Co content is only ~11.5%. Most importantly, the uniformity in *inter*-layer distances between the metal oxide layers is maintained during delithiation in case of Na_NCAs even in the case of lowest Co content (Na_NCA_IV). The uniformity in the oxide layers during delithiation is an essential feature in order to resist the structural collapse and it is clear from the AIMD simulations that Na_NCA_IV (at 100%, 50% and 0% Li concentration) maintains the structural uniformity during the course of simulation.

The combined DFT and AIMD investigation advocate the use of NCA materials with lowest possible Co content and the potential of incorporating sodium in order to prevent the collapse of metal oxide layers and resist Li/Ni cation mixing. Limited use of Co is one of the most critical concerns for the ever-growing battery industry due to its environmental hazards, high cost, and undersupply. Thus, our detailed first principle calculations on NCA type cathode materials depicts a thorough systematic investigation on the broad range of varying Co content. Our study provides an essential atomistic level fundamental insight into the efficient design of cathode materials for next-generation batteries.

**Supporting Information**
Detailed calculations of the cost-effectiveness with reduced cobalt concentration; Tables for (i) Comparison of structural parameters (inter atomic distances) of **NCA_I** and **NCA_IV** for fully lithiated, 50% lithiated and completely delithiated states; (ii) Comparison of structural parameters (inter atomic distances) of **Na_NCA_I** and **Na_NCA_IV** for fully lithiated, 50% lithiated and completely delithiated states; and (iii) Cell volume change of NCA_Is, NCA_IVs during the lithiation/delithiation process.

**Acknowledgement**
DD acknowledges NJIT for the faculty start-up package. We thank Prof. Siva Nadimpalli of NJIT for thoughtful comments during the manuscript preparation. We are grateful to the High-Performance Computing (HPC) facilities managed by Academic and Research Computing Systems (ARCS) in the Department of Information Services and Technology (IST) of the New Jersey Institute of Technology (NJIT). Some computations were performed on Kong.njit.edu HPC cluster, managed by ARCS. We acknowledge the support of the Extreme Science and Engineering Discovery Environment (XSEDE) for providing us their computational facilities (Start Up Allocation - DMR170065 &

# References

1. Obrovac, M.; Mao, O.; Dahn, J., Structure and electrochemistry of LiMO 2 (M= Ti, Mn, Fe, Co, Ni) prepared by mechanochemical synthesis. *Solid State Ionics* **1998,** *112* (1), 9-19.
2. Mizushima, K.; Jones, P.; Wiseman, P.; Goodenough, J. B., LixCoO2 (0< x<-1): A new cathode material for batteries of high energy density. *Materials Research Bulletin* **1980,** *15* (6), 783-789.
3. Zaghib, K.; Julien, C.; Prakash, J. In *New Trends in Intercalation Compounds for Energy Storage and Conversion: Proceedings of the International Symposium*, The Electrochemical Society: 2003.
4. Du Pasquier, A.; Plitz, I.; Menocal, S.; Amatucci, G., A comparative study of Li-ion battery, supercapacitor and nonaqueous asymmetric hybrid devices for automotive applications. *Journal of Power Sources* **2003,** *115* (1), 171-178.
5. Tollefson, J., Charging up the future. *Nature* **2008,** *456* (7221), 436.
6. Dahn, J.; Fuller, E.; Obrovac, M.; Von Sacken, U., Thermal stability of LixCoO2, LixNiO2 and λ-MnO2 and consequences for the safety of Li-ion cells. *Solid State Ionics* **1994,** *69* (3-4), 265-270.
7. Orendorff, C. J.; Doughty, D. H., Lithium ion battery safety. *The Electrochemical Society Interface* **2012,** *21* (2), 35-35.
8. Etacheri, V.; Marom, R.; Elazari, R.; Salitra, G.; Aurbach, D., Challenges in the development of advanced Li-ion batteries: a review. *Energy & Environmental Science* **2011,** *4* (9), 3243-3262.
9. Bruce, P. G., Solid-state chemistry of lithium power sources. *Chemical Communications* **1997,** (19), 1817-1824.
10. Choi, N. S.; Chen, Z.; Freunberger, S. A.; Ji, X.; Sun, Y. K.; Amine, K.; Yushin, G.; Nazar, L. F.; Cho, J.; Bruce, P. G., Challenges facing lithium batteries and electrical double-layer capacitors. *Angewandte Chemie International Edition* **2012,** *51* (40), 9994-10024.
11. Liu, W.; Oh, P.; Liu, X.; Lee, M. J.; Cho, W.; Chae, S.; Kim, Y.; Cho, J., Nickel-Rich Layered Lithium Transition-Metal Oxide for High-Energy Lithium-Ion Batteries. *Angewandte Chemie International Edition* **2015,** *54* (15), 4440-4457.
12. Lee, K. K.; Kim, K. B., Electrochemical and Structural Characterization of LiNi1− y Co y O 2 (0≤ y≤ 0.2) Positive Electrodes during Initial Cycling. *Journal of The Electrochemical Society* **2000,** *147* (5), 1709-1717.
13. Rougier, A.; Gravereau, P.; Delmas, C., Optimization of the Composition of the Li1− z Ni1+ z O 2 Electrode Materials: Structural, Magnetic, and Electrochemical Studies. *Journal of The Electrochemical Society* **1996,** *143* (4), 1168-1175.
14. Chikkannanavar, S. B.; Bernardi, D. M.; Liu, L., A review of blended cathode materials for use in Li-ion batteries. *Journal of Power Sources* **2014,** *248*, 91-100.

15. Lin, F.; Markus, I. M.; Nordlund, D.; Weng, T.-C.; Asta, M. D.; Xin, H. L.; Doeff, M. M., Surface reconstruction and chemical evolution of stoichiometric layered cathode materials for lithium-ion batteries. *Nature Communications* **2014,** *5*, 3529.
16. Kam, K. C.; Mehta, A.; Heron, J. T.; Doeff, M. M., Electrochemical and physical properties of Ti-substituted layered nickel manganese cobalt oxide (NMC) cathode materials. *Journal of The Electrochemical Society* **2012,** *159* (8), A1383-A1392.
17. Lee, M.-J.; Noh, M.; Park, M.-H.; Jo, M.; Kim, H.; Nam, H.; Cho, J., The role of nanoscale-range vanadium treatment in LiNi 0.8 Co 0.15 Al 0.05 O 2 cathode materials for Li-ion batteries at elevated temperatures. *Journal of Materials Chemistry A* **2015,** *3* (25), 13453-13460.
18. Kalyani, P.; Kalaiselvi, N., Various aspects of LiNiO 2 chemistry: A review. *Science and Technology of Advanced Materials* **2005,** *6* (6), 689-703.
19. Chen, C.; Liu, J.; Stoll, M.; Henriksen, G.; Vissers, D.; Amine, K., Aluminum-doped lithium nickel cobalt oxide electrodes for high-power lithium-ion batteries. *Journal of power Sources* **2004,** *128* (2), 278-285.
20. Nitta, N.; Wu, F.; Lee, J. T.; Yushin, G., Li-ion battery materials: present and future. *Materials today* **2015,** *18* (5), 252-264.
21. Cai, L.; Liu, Z.; An, K.; Liang, C., Probing Li-Ni Cation Disorder in Li1− xNi1+ x− yAlyO2 Cathode Materials by Neutron Diffraction. *Journal of The Electrochemical Society* **2012,** *159* (7), A924-A928.
22. Muto, S.; Tatsumi, K.; Kojima, Y.; Oka, H.; Kondo, H.; Horibuchi, K.; Ukyo, Y., Effect of Mg-doping on the degradation of LiNiO 2-based cathode materials by combined spectroscopic methods. *Journal of Power Sources* **2012,** *205*, 449-455.
23. Huang, B.; Li, X.; Wang, Z.; Guo, H.; Xiong, X., Synthesis of Mg-doped LiNi 0.8 Co 0.15 Al 0.05 O 2 oxide and its electrochemical behavior in high-voltage lithium-ion batteries. *Ceramics International* **2014,** *40* (8), 13223-13230.
24. Liu, Z.; Fan, Y.; Peng, W.; Wang, Z.; Guo, H.; Li, X.; Wang, J., Mechanical activation assisted soft chemical synthesis of Na-doped lithium vanadium fluorophosphates with improved lithium storage properties. *Ceramics International* **2015,** *41* (3), 4267-4271.
25. Park, S.-H.; Shin, S.-S.; Sun, Y.-K., The effects of Na doping on performance of layered Li 1.1− x Na x [Ni 0.2 Co 0.3 Mn 0.4] O 2 materials for lithium secondary batteries. *Materials chemistry and physics* **2006,** *95* (2), 218-221.
26. Hua, W.; Zhang, J.; Zheng, Z.; Liu, W.; Peng, X.; Guo, X.-D.; Zhong, B.; Wang, Y.-J.; Wang, X., Na-doped Ni-rich LiNi 0.5 Co 0.2 Mn 0.3 O 2 cathode material with both high rate capability and high tap density for lithium ion batteries. *Dalton Transactions* **2014,** *43* (39), 14824-14832.
27. Xie, H.; Du, K.; Hu, G.; Peng, Z.; Cao, Y., The role of sodium in LiNi0. 8Co0. 15Al0. 05O2 cathode material and its electrochemical behaviors. *The Journal of Physical Chemistry C* **2016,** *120* (6), 3235-3241.
28. Kang, K.; Meng, Y. S.; Bréger, J.; Grey, C. P.; Ceder, G., Electrodes with high power and high capacity for rechargeable lithium batteries. *Science* **2006,** *311* (5763), 977-980.
29. Lin, Y.-M.; Abel, P. R.; Gupta, A.; Goodenough, J. B.; Heller, A.; Mullins, C. B., Sn–Cu nanocomposite anodes for rechargeable sodium-ion batteries. *ACS applied materials & interfaces* **2013,** *5* (17), 8273-8277.

30. Seyfried, W.; Janecky, D.; Mottl, M., Alteration of the oceanic crust: implications for geochemical cycles of lithium and boron. *Geochimica et Cosmochimica Acta* **1984,** *48* (3), 557-569.
31. Kim, S. W.; Seo, D. H.; Ma, X.; Ceder, G.; Kang, K., Electrode materials for rechargeable sodium-ion batteries: potential alternatives to current lithium-ion batteries. *Advanced Energy Materials* **2012,** *2* (7), 710-721.
32. Kraytsberg, A.; Ein-Eli, Y., Higher, Stronger, Better… A Review of 5 Volt Cathode Materials for Advanced Lithium-Ion Batteries. *Advanced Energy Materials* **2012,** *2* (8), 922-939.
33. Institute, T. C. d., http://thecdi.com/.
34. Petersen, J., Tesla's Evolving Cobalt Nightmare. *Seeking Alpha*

http://seekingalpha.com/article/4027400-teslas-evolving-cobalt-nightmare **2016,** (4027400).
35. Roberts, S.; Gunn, G., Cobalt. In *Critical Metals Handbook*, John Wiley & Sons: 2014; pp 122-149.
36. Hwang, S.; Chang, W.; Kim, S. M.; Su, D.; Kim, D. H.; Lee, J. Y.; Chung, K. Y.; Stach, E. A., Investigation of Changes in the Surface Structure of $Li_x Ni_{0.8}Co_{0.15}Al_{0.05}O_2$ Cathode Materials Induced by the Initial Charge. *Chemistry of Materials* **2014,** *26* (2), 1084-1092.
37. Jo, M.; Noh, M.; Oh, P.; Kim, Y.; Cho, J., A New High Power $LiNi_{0.81}Co_{0.1}Al_{0.09}O_2$ Cathode Material for Lithium-Ion Batteries. *Advanced Energy Materials* **2014,** *4* (13).
38. Du, K.; Xie, H.; Hu, G.; Peng, Z.; Cao, Y.; Yu, F., Enhancing the Thermal and Upper Voltage Performance of Ni-Rich Cathode Material by a Homogeneous and Facile Coating Method: Spray-Drying Coating with Nano-$Al_2O_3$. *ACS applied materials & interfaces* **2016,** *8* (27), 17713-17720.
39. Casals, L. C.; García, B. A. In *A review of the complexities of applying second life electric car batteries on energy businesses*, Energy Syst. Conf, 2014.
40. SONY, A Dream Comes True: The Lithium-Ion Rechargeable Battery. http://www.sony.net/SonyInfo/CorporateInfo/History/SonyHistory/2-13.html - block3 *Part II; Chapter13 Recognized as an International Standard*.
41. Nagaura, T.; Tozawa, K., Progress in batteries and solar cells. *JEC Press* **1990,** *9*, 209.
42. Winter, M.; Moeller, K.; Besenhard, J.; Nazri, G.; Pistoia, G., Lithium Batteries: Science and Technology. *Nazri, G.-A* **2004**, 148.
43. Dixit, M.; Markovsky, B.; Schipper, F.; Aurbach, D.; Major, D. T., Origin of Structural Degradation During Cycling and Low Thermal Stability of Ni-Rich Layered Transition Metal-Based Electrode Materials. *The Journal of Physical Chemistry C* **2017,** *121* (41), 22628-22636.
44. Aydinol, M.; Kohan, A.; Ceder, G.; Cho, K.; Joannopoulos, J., Ab initio study of lithium intercalation in metal oxides and metal dichalcogenides. *Physical Review B* **1997,** *56* (3), 1354.
45. Zhi, M.; Xiang, C.; Li, J.; Li, M.; Wu, N., Nanostructured carbon–metal oxide composite electrodes for supercapacitors: a review. *Nanoscale* **2013,** *5* (1), 72-88.
46. Kresse, G.; Furthmüller, J., Efficient iterative schemes for ab initio total-energy calculations using a plane-wave basis set. *Physical review B* **1996,** *54* (16), 11169.

47. Blöchl, P. E., Projector augmented-wave method. *Physical review B* **1994,** *50* (24), 17953.
48. Kresse, G.; Joubert, D., From ultrasoft pseudopotentials to the projector augmented-wave method. *Physical Review B* **1999,** *59* (3), 1758.
49. Perdew, J. P.; Burke, K.; Wang, Y., Generalized gradient approximation for the exchange-correlation hole of a many-electron system. *Physical Review B* **1996,** *54* (23), 16533.
50. Hirano, A.; Kanno, R.; Kawamoto, Y.; Takeda, Y.; Yamaura, K.; Takano, M.; Ohyama, K.; Ohashi, M.; Yamaguchi, Y., Relationship between non-stoichiometry and physical properties in LiNiO2. *Solid State Ionics* **1995,** *78* (1), 123-131.
51. Bergerhoff, G.; Hundt, R.; Sievers, R.; Brown, I. D., The inorganic crystal structure data base. *Journal of Chemical Information and Computer Sciences* **1983,** *23* (2), 66-69.
52. Aydinol, M.; Ceder, G., First-Principles Prediction of Insertion Potentials in Li-Mn Oxides for Secondary Li Batteries. *Journal of the Electrochemical Society* **1997,** *144* (11), 3832-3835.
53. Stournara, M. E.; Shenoy, V. B., Enhanced Li capacity at high lithiation potentials in graphene oxide. *Journal of Power Sources* **2011,** *196* (13), 5697-5703.
54. Zhou, F.; Cococcioni, M.; Kang, K.; Ceder, G., The Li intercalation potential of LiMPO4 and LiMSiO4 olivines with M=Fe, Mn, Co, Ni. *Electrochemistry Communications* **2004,** *6* (11), 1144-1148.
55. Martha, S. K.; Haik, O.; Zinigrad, E.; Exnar, I.; Drezen, T.; Miners, J. H.; Aurbach, D., On the thermal stability of olivine cathode materials for lithium-ion batteries. *Journal of the Electrochemical Society* **2011,** *158* (10), A1115-A1122.
56. Liu, C.; Neale, Z. G.; Cao, G., Understanding electrochemical potentials of cathode materials in rechargeable batteries. *Materials Today* **2016,** *19* (2), 109-123.
57. Bak, S.-M.; Nam, K.-W.; Chang, W.; Yu, X.; Hu, E.; Hwang, S.; Stach, E. A.; Kim, K.-B.; Chung, K. Y.; Yang, X.-Q., Correlating Structural Changes and Gas Evolution during the Thermal Decomposition of Charged Li x Ni0. 8Co0. 15Al0. 05O2 Cathode Materials. *Chemistry of Materials* **2013,** *25* (3), 337-351.
58. Nam, K.-W.; Bak, S.-M.; Hu, E.; Yu, X.; Zhou, Y.; Wang, X.; Wu, L.; Zhu, Y.; Chung, K.-Y.; Yang, X.-Q., Combining In Situ Synchrotron X-Ray Diffraction and Absorption Techniques with Transmission Electron Microscopy to Study the Origin of Thermal Instability in Overcharged Cathode Materials for Lithium-Ion Batteries. *Advanced Functional Materials* **2013,** *23* (8), 1047-1063.
59. Wang, L. P.; Yu, L.; Wang, X.; Srinivasan, M.; Xu, Z. J., Recent developments in electrode materials for sodium-ion batteries. *Journal of Materials Chemistry A* **2015,** *3* (18), 9353-9378.
60. Ong, S. P.; Chevrier, V. L.; Hautier, G.; Jain, A.; Moore, C.; Kim, S.; Ma, X.; Ceder, G., Voltage, stability and diffusion barrier differences between sodium-ion and lithium-ion intercalation materials. *Energy & Environmental Science* **2011,** *4* (9), 3680-3688.
61. Whittingham, M. S., Lithium batteries and cathode materials. *Chemical reviews* **2004,** *104* (10), 4271-4302.

62. Xiong, F.; Yan, H. J.; Chen, Y.; Xu, B.; Le, J. X.; Ouyang, C. Y., The Atomic and Electronic Structure Changes Upon Delithiation of LiCoO2: From First Principles Calculations. *International Journal of ELECTROCHEMICAL SCIENCE* **2012,** *7* (10).

63. Li, W.; Reimers, J. N.; Dahn, J. R., In situ x-ray diffraction and electrochemical studies of Li1−xNiO2. *Solid State Ionics* **1993,** *67* (1–2), 123-130.

64. Amatucci, G. G.; Tarascon, J. M.; Klein, L. C., CoO2, The End Member of the Li x CoO2 Solid Solution. *Journal of The Electrochemical Society* **1996,** *143* (3), 1114-1123.

65. Shibuya, M.; Nishina, T.; Matsue, T.; Uchida, I., In situ conductivity measurements of LiCoO2 film during lithium insertion/extraction by using interdigitated microarray electrodes. *Journal of The Electrochemical Society* **1996,** *143* (10), 3157-3160.

66. Dutta, G.; Manthiram, A.; Goodenough, J.; Grenier, J.-C., Chemical synthesis and properties of Li1− δ− xNi1+ δO2 and Li [Ni2] O4. *Journal of Solid State Chemistry* **1992,** *96* (1), 123-131.

67. Kim, H.; Hong, J.; Park, K.-Y.; Kim, H.; Kim, S.-W.; Kang, K., Aqueous rechargeable Li and Na ion batteries. *Chemical reviews* **2014,** *114* (23), 11788-11827.

68. Palomares, V.; Serras, P.; Villaluenga, I.; Hueso, K. B.; Carretero-González, J.; Rojo, T., Na-ion batteries, recent advances and present challenges to become low cost energy storage systems. *Energy & Environmental Science* **2012,** *5* (3), 5884-5901.

69. Liu, J.; Zhang, J. G.; Yang, Z.; Lemmon, J. P.; Imhoff, C.; Graff, G. L.; Li, L.; Hu, J.; Wang, C.; Xiao, J., Materials science and materials chemistry for large scale electrochemical energy storage: from transportation to electrical grid. *Advanced Functional Materials* **2013,** *23* (8), 929-946.

70. Mortazavi, M.; Deng, J.; Shenoy, V. B.; Medhekar, N. V., Elastic softening of alloy negative electrodes for Na-ion batteries. *Journal of Power Sources* **2013,** *225*, 207-214.

71. Zhu, H.; Jia, Z.; Chen, Y.; Weadock, N.; Wan, J.; Vaaland, O.; Han, X.; Li, T.; Hu, L., Tin anode for sodium-ion batteries using natural wood fiber as a mechanical buffer and electrolyte reservoir. *Nano letters* **2013,** *13* (7), 3093-3100.

72. Zhang, J.; Lee, J., A review on prognostics and health monitoring of Li-ion battery. *Journal of Power Sources* **2011,** *196* (15), 6007-6014.

# Supporting Information

# Effect of Cobalt Content on the Electrochemical Properties and Structural Stability of NCA Type Cathode Materials

Kamalika Ghatak[1], Swastik Basu[2], Tridip Das[3], Hemant Kumar[4], Dibakar Datta[1,*]

[1] Department of Mechanical and Industrial Engineering, Newark College of Engineering, New Jersey Institute of Technology (NJIT), Newark, NJ 07102, USA

[2] Department of Mechanical, Aerospace, and Nuclear Engineering, Rensselaer Polytechnic Institute, Troy, NY 12180, USA

[3] Department of Chemical Engineering and Materials Science, Michigan State University, East Lansing, MI 48824, USA

[4] Department of Materials Science and Engineering, University of Pennsylvania, Philadelphia, PA 19104, USA

**Corresponding Author:**

Dibakar Datta; Phone: 973-596-3647; Email: dibakar.datta@njit.edu

## Section I. Cost-effectiveness with reduced cobalt concentration

We can estimate the overall cost effectiveness as follow:

Cobalt cost/lb (15$^{th}$ March, 2018)[4] = 39.58 USD

Nickel cost/lb (15$^{th}$ March, 2018)[5] = 6.21 USD

Cost reduction due to decrease in cobalt concentration = 29.68 USD/lb

Cost increment due to increase in nickel concentration = 0.98 USD/lb

The overall cost saving = 28.7 USD/lb

Most importantly, the cost of Cobalt increased in significant extent during the last year itself. The increment between January 2017 (14.2 USD/lb) and March 2018 (39.58 USD/lb) is ~25.38 USD/lb. An overall 75% reduction of Co concentration will not only lead to huge cost saving for the battery production, but it also implements an additional environmental benefit due to the reduction of overall toxicity. Moreover, NCA_IV and Na_NCA_IV, containing mostly Ni, are perfect agents for economical usage due to the cheaper cost and higher capacity of Ni as compared to the Co.[6-8].

**Table S1.** Comparison of structural parameters (inter atomic distances) of **NCA_I** and **NCA_IV** for fully lithiated, 50% lithiated and completely delithiated states.

| System | avg. bond dist. (Ni-O) in Å | avg. bond dist. (Co-O) in Å | avg. bond dist. (Al-O) in Å | avg. intra layer bond dist. (Ni-Ni) in Å | avg. inter layer dist. (Ni-Ni) in Å |
|---|---|---|---|---|---|
| NCA_I_Li$_{24}$ | 2.003 | 1.931 | 1.932 | 2.879 | 5.053 |
| NCA_I_Li$_{12}$ | 1.912 | 1.903 | 1.925 | 2.835 | 5.153 |
| NCA_I_Li$_{0}$ | 1.866 | 1.882 | 1.899 | 2.831 | 5.228 |
| NCA_IV_Li$_{24}$ | 1.948 | 2.001 | 1.942 | 2.867 | 5.039 |
| NCA_IV_Li$_{12}$ | 1.894 | 1.984 | 1.924 | 2.834 | 5.155 |
| NCA_IV_Li$_{0}$ | 1.856 | 1.874 | 1.896 | 2.780 | 5.216 |

**Table S2.** Comparison of structural parameters (inter atomic distances) of **Na_NCA_I** and **Na_NCA_IV** for fully lithiated, 50% lithiated and completely delithiated states.

| System | avg. bond dist. (Ni-O) in Å | avg. bond dist. (Co-O) in Å | avg. bond dist. (Al-O) in Å | avg. intra layer bond dist. (Ni-Ni) in Å | avg. inter layer dist. (Ni-Ni) in Å |
|---|---|---|---|---|---|
| Na_NCA_I_Li$_{22}$ | 1.970 | 1.887 | 1.948 | 2.887 | 5.109 |
| Na_NCA_I_Li$_{10}$ | 1.898 | 1.881 | 1.937 | 2.867 | 5.231 |
| Na_NCA_I_Li$_0$ | 1.894 | 1.873 | 1.917 | 2.871 | 5.610 |
| Na_NCA_IV_Li$_{22}$ | 1.975 | 1.944 | 1.942 | 2.898 | 5.113 |
| Na_NCA_IV_Li$_{10}$ | 1.906 | 1.892 | 1.931 | 2.824 | 5.347 |
| Na_NCA_IV_Li$_0$ | 1.886 | 1.885 | 1.910 | 2.787 | 5.893 |

**Table S3.** Cell volume change of NCA_Is, NCA_IVs during the lithiation/delithiation process.

| System | Volume Å$^3$ |
|---|---|
| NCA_I_LI0 | 822.124 |
| NCA_I_LI12 | 816.688 |
| NCA_I_LI24 | 819.508 |
| NCA_IV_LI0 | 795.612 |
| NCA_IV_LI12 | 815.183 |
| NCA_IV_LI24 | 818.368 |
| Na_NCA_I_LI0 | 876.721 |
| Na_NCA_I_LI10 | 837.450 |
| Na_NCA_I_LI22 | 837.058 |
| Na_NCA_IV_LI0 | 880.819 |
| Na_NCA_IV_LI10 | 836.747 |
| Na_NCA_IV_LI22 | 830.356 |

**References**

1. Jain, A.; Ong, S. P.; Hautier, G.; Chen, W.; Richards, W. D.; Dacek, S.; Cholia, S.; Gunter, D.; Skinner, D.; Ceder, G. Commentary: The Materials Project: A Materials Genome Approach to Accelerating Materials Innovation. *Apl Materials* **2013**, *1* (1), 011002.
2. 765279, M. p. m.-. *https://materialsproject.org/materials/mp-765279/*.
3. Momma, K.; Izumi, F. Vesta: A Three-Dimensional Visualization System for Electronic and Structural Analysis. *Journal of Applied Crystallography* **2008**, *41* (3), 653-658.
4. Cobalt Prices. http://www.infomine.com/investment/metal-prices/cobalt/1-year/.
5. Nickel Prices. http://www.infomine.com/investment/metal-prices/nickel/1-year/.
6. Sakamoto, K.; Hirayama, M.; Sonoyama, N.; Mori, D.; Yamada, A.; Tamura, K.; Mizuki, J. i.; Kanno, R. Surface Structure of Lini0. 8co0. 2o2: A New Experimental Technique Using in Situ X-Ray Diffraction and Two-Dimensional Epitaxial Film Electrodes. *Chemistry of Materials* **2009**, *21* (13), 2632-2640.
7. Baskaran, R.; Kuwata, N.; Kamishima, O.; Kawamura, J.; Selvasekarapandian, S. Structural and Electrochemical Studies on Thin Film Lini 0.8 Co 0.2 O 2 by Pld for Micro Battery. *Solid State Ionics* **2009**, *180* (6), 636-643.
8. Li, D.; Peng, Z.; Ren, H.; Guo, W.; Zhou, Y. Synthesis and Characterization of Lini 1− X Co X O 2 for Lithium Batteries by a Novel Method. *Materials Chemistry and Physics* **2008**, *107* (1), 171-176.